# Second harmonic generation in anisotropic stratified media: A generalization of the Berreman method and its application to photonic materials.

J. ORTEGA, C. L. FOLCIA AND J. ETXEBARRIA*

*Department of Physics, Faculty of Science and Technology, UPV/EHU, Bilbao, Spain*

*\* j.etxebarria@ehu.eus*

## Abstract

We have developed a numerical method for calculating the second harmonic generation (SHG) generated by an anisotropic material whose optical properties present an arbitrary modulation in one dimension. The method is based on the Berreman 4x4 matrix formalism, which is generalized to include nonlinear optical phenomena. It can be used under oblique incidences of the input beam, and is valid even when the SHG frequency is close to photonic bands, where the usual slowly-varying-amplitude approximation breaks down. As an example of application, we have studied the SHG performance of ferroelectric and helielectric nematic liquid crystals. The latter present a helicoidal structure that can be distorted under electric field. In the different tests of the method we have analyzed the conditions for the most efficient SHG, and compared with previous results in the case there were any. The obtained results indicate that the present procedure may contribute to improve the structural design and enlarge the variety of nonlinear optical materials for their application in optical devices.

## 1. Introduction

The process of second-harmonic generation (SHG) in optics is a particular case of three wave mixing that allows two photons of frequency $\omega$ to be converted into a photon of frequency $2\omega$ [1, 2]. The effect is very interesting from a technological point of view and can be used, for example, in laser technologies to obtain multi-line emission systems. However, usually the SHG conversion is very poor in homogeneous crystals due to the short coherence length that the SHG process typically presents. The SHG coherence length increases significantly when the so-called phase matching (PM) condition occurs. In this case, the phase mismatch between the fundamental and second-harmonic (SH) waves is zero. In homogeneous materials, this effect can occur along particular directions and light polarizations and, in practice, the effect gives rise to large coherence lengths, which has been traditionally used for the construction of optical devices [3-5].

The SHG performance can be significantly improved by using optical systems with modulated dielectric properties. The most common structures of this kind are stratified optical media with a modulation in one dimension (1D). In these systems, new possibilities of PM occur since the wave vector mismatch between the fundamental and SH fields can be compensated with the wave vector of the dielectric modulation [6]. In general, the periodic systems present photonic properties and exhibit a spectral pattern of photonic band gaps (PBGs) that depends on the particular profile of the modulation of the dielectric tensor. Especially interesting are the PM situations that appear when the wavelength of the lights involved in the SHG process coincide with the edge of the PBGs. In these cases, the electromagnetic wave is enhanced by resonance, giving rise to a significant increase of the SHG efficiency [7-13]. The process is especially favorable when the wavelengths of both the fundamental and SH lights coincide simultaneously with edges of the photonic system [11]. Under these circumstances the SHG signal of the optical system is dramatically enhanced. However, this process is only possible in some modulated systems that, in most cases, are artificially manufactured. Typically, they consist of multilayer periodic structures made of different dielectric materials that are SHG active. In this kind of systems, the reflectance spectrum presents a set of PBGs whose frequencies occur at multiples of a fundamental one. In order to obtain the desired coincidence of the fundamental and SH frequencies at edges of the PBGs, the periodicity of the layered structure must compensate the normal dispersion of the refractive indices of the materials [12]. In addition, the SHG performance also depends on the coupling of the polarization of the fundamental wave with the local second-order susceptibility tensor. In general, the calculation of the SHG intensity generated by a photonic material is cumbersome and only a few examples of modulated structures have been solved so far. In some cases, the studies are restricted to the so-called slowly-varying-amplitude approximation (SVAA) for the SHG field, which exclude the SHG description at the band edges [14]. In other cases, the problem is solved only for optical systems for which there exist analytical expressions for the polarization eigenstates of the fundamental and SHG fields [7,8]. The problem has also been addressed for isotropic systems modulated in an arbitrary way [9-13]. However, the general case of anisotropic systems with arbitrary modulation and at oblique incidences has not been solved to the best of our knowledge. The development of a tool for studying this problem is very interesting to explore other possibilities of high-performance SHG materials.

In this work, we have developed a numerical method for calculating the SHG field generated by an anisotropic material whose dielectric tensor presents an arbitrary modulation in 1D, even for input lights at oblique incidence. The method is a generalization of the Berreman formalism [15,16] that is commonly used to study linear optical processes in inhomogeneous materials. As an output, we obtain the intensity of the SHG wave transmitted and reflected by the material in absolute units. The only assumption of the method is that the power depletion of the incident beam can be neglected. The procedure can be applied even in the cases where the amplitude of the SHG electric field varies rapidly along the propagation direction. This point is important because the SVAA is violated in situations where the SHG is especially efficient, which present an evident practical interest.

As an example of validation and application of our calculation method, we have studied several optical systems based on ferroelectric nematic liquid crystals ($N_F$) [17-27]. The main features of this type of materials will be discussed later. The reason for this election is twofold. On the one hand, these organic materials can exhibit high SHG responses since both the structure of the molecular constituents and the way the molecules are arranged in the phase are very favorable for SHG purposes. On the other hand, in the chiral variants of this type of phases the molecules self-assemble naturally in helicoidal organizations. This is an important advantage with respect to artificial multilayer systems whose fabrication is difficult and expensive. In addition, the structure of the chiral ferroelectric nematic ($N_F$*) phases can be modified by low electric fields, giving rise to interesting spectral patterns of the PBGs [28-31].

This work is structured as follows: In the next section we will describe the basics of the calculation procedure, next we will demonstrate the validity of the method in several examples for which there are available theories. Finally, we will study the SHG performance in the $N_F$* phase distorted by electric fields. At the end of the paper some conclusions will be drawn.

## 2. Basics of the calculation of the SHG efficiency.

First of all, we will make some comments about the well-known (linear) Berreman method. This 4x4 matrix technique allows us to calculate numerically the light transmitted and reflected by an anisotropic stratified medium with an arbitrary modulation of the dielectric tensor. The method can be also applied to continuously modulated systems by approximating them to discrete stratified media. The method uses the so-called Berreman vectors defined as:

$$\Psi = \begin{pmatrix} E_x \\ H_y \\ E_y \\ -H_x \end{pmatrix} \quad (1)$$

where $E_{x,y}$ and $H_{x,y}$ are the electric and magnetic fields in a reference frame where $z$ is perpendicular to the layers and $(x,z)$ define the plane of incidence. The Berreman vectors are continuous through the planes separating two consecutive layers of the medium. Every layer $l$ ($l$=1,…,$n$) of the medium is represented by a transfer matrix that depends on the local dielectric tensor, the frequency of the light and the angle of incidence. Therefore, given an input Berreman vector, the output vector through the whole system is obtained by applying subsequently the whole set of transfer matrices corresponding to the different layers. Moreover, under the Berreman formalism, it is possible to calculate the electric and magnetic fields at every inner layer provided we know the corresponding input vector.

Let us now consider an inhomogeneous anisotropic nonlinear medium illuminated by a fundamental beam of frequency $\omega$. To calculate the SHG electric field generated at a particular layer of the stratified medium we will proceed as follows:

1. First we calculate the fundamental electric field at that layer, corresponding to a given input light. For an input Berreman vector $\mathbf{\psi}_{input}^{\omega}$, the Berreman vector at the $l$-th layer $\mathbf{\psi}^{\omega}(l)$ is given by:

$$\mathbf{\psi}^{\omega}(l) = U^{\omega}(l)...U^{\omega}(2)U^{\omega}(1)\mathbf{\psi}_{input}^{\omega} \qquad (2)$$

where $U^{\omega}(l)$ is the transfer matrix of the $l$th layer for frequency $\omega$, which can be computed following standard methods in the Berreman procedure [15]. As a result of this calculation we obtain the electric field of frequency $\omega$ at any point of the medium $E_j^{\omega}(l)$ ($j$=$x,y,z$; $l$=1,…,$n$)

2. Now if the medium has a second-order susceptibility tensor $d_{ijk}(l)$ at layer $l$, there is a polarization $P_i(l)$ induced by the fundamental electric field which, up to second order, is given by the expression:

$$P_i(l) = \varepsilon_0 \chi_{ij}(l) E_j^{\omega}(l) + 2\varepsilon_0 d_{ijk}(l) E_j^{\omega}(l) E_k^{\omega}(l) \qquad (3)$$

where $\chi_{ij}(l)$ is the linear susceptibility at layer $l$, and $\varepsilon_0$ the vacuum permittivity. Eq. (3) can be understood as if the medium now presented a dielectric tensor $\varepsilon'_{ij}(l)$ distorted by the fundamental electric field and given by:

$$\varepsilon'_{ij}(l) = \varepsilon_{ij}(l) + 2 d_{ijk}(l) E_k^{\omega}(l) \qquad (4)$$

where $\varepsilon_{ij}(l) = \delta_{ij} + \chi_{ij}(l)$ is the dielectric tensor of the non-distorted material and $\delta_{ij}$ the Kronecker delta. By using $\varepsilon'_{ij}(l)$, we can re-calculate a new transfer matrix $U'^{\omega}(l)$ for the $l$th layer of the distorted structure. The Berreman vector corresponding to the SH field generated at this layer ($\mathbf{\psi}^{2\omega}(l)$) can be obtained by means of the expression:

$$\mathbf{\psi}^{2\omega}(l) = \left[U'^{\omega}(l) - U^{\omega}(l)\right]\mathbf{\psi}^{\omega}(l) \qquad (5)$$

where $\mathbf{\psi}^{\omega}(l)$ is the Berreman vector of the fundamental input light at the $l$th layer. On writing Eq. (5) we state that the excess of electric and magnetic fields, generated in the distorted material (with respect to the undistorted original medium), are the fields corresponding to the SHG wave, which is generated as a consequence of the second-order nonlinear response of the material.

Note, however, that the fields directly calculated from Eq. (5) have still to be corrected in order to represent properly the SHG fields, because the optical properties of the surrounding medium can remarkably alter the SH fields that finally give rise to the SHG signal outside the material. An example will clarify this point. Suppose a case in which the frequency $2\omega$ is inside a PBG of the structure. In this situation, the amplitude of some normal modes will be heavily penalized in their contribution to the SHG field. The modes with small contributions will be precisely those that cannot propagate in the gap and are strongly attenuated inside the sample. A similar circumstance occurs when studying the fluorescent emission of a dye molecule immersed in a photonic medium: depending on the available density of photonic states the resulting fluorescence is greatly altered with respect to the same emission in vacuum [32-36]. The algorithm to calculate the true Berreman vector of the SH field generated in layer $l$ starting from $\boldsymbol{\psi}^{2\omega}(l)$ is explained in the next point.

3. To calculate the correction to the $\boldsymbol{\psi}^{2\omega}(l)$ vector we will begin by expressing it as a linear combination of 4 linearly independent Berreman vectors. For each layer $l$, these 4 vectors $\mathbf{v}_k(l)$ ($k$=1,…4) will be constructed from the Berreman vectors $\mathbf{v}_k$ associated to the polarizations of the four transmission eigenstates of the sample with unit intensity that propagate forward and backward. Vectors $\mathbf{v}_k$ can be easily determined following well-established procedures in the Berreman formalism [33,36], whereas $\mathbf{v}_k(l)$ are deduced from them using the expressions

$$\mathbf{v}_k(l) = U^{2\omega}(l)…U^{2\omega}(2)U^{2\omega}(1)\mathbf{v}_k \qquad (6)$$

where $U^{2\omega}(l)$ is the transfer matrix for layer $l$ and frequency $2\omega$.

It should be noted that although, by construction, the vectors $\mathbf{v}_k$ are normalized in the sense that the intensities associated to them are equal to unity, the vectors $\mathbf{v}_k(l)$ are not, and their intensities can be very different from each other.

In general, it can be shown that the time-averaged $z$ component of the Poynting vector $S_z$ for the light wave associated to a Berreman vector $\mathbf{v}$ is given by [16]

$$S_z = \frac{c\varepsilon_0}{4}\mathbf{v}^\dagger.M.\mathbf{v} \qquad (7)$$

where $c$ is the speed of light in vacuum, and

$$M = \begin{pmatrix} 0 & 1 & 0 & 0 \\ 1 & 0 & 0 & 0 \\ 0 & 0 & 0 & 1 \\ 0 & 0 & 1 & 0 \end{pmatrix} \qquad (8)$$

The matrix $M$ can then be understood as a metric tensor which allows us to define something analogous to a scalar product of a pair of Berreman vectors $\mathbf{v}_1$ and $\mathbf{v}_2$ ($\mathbf{v}_1^\dagger.M.\mathbf{v}_2$), and a norm $\|\mathbf{v}\| = \sqrt{|\mathbf{v}^\dagger.M.\mathbf{v}|}$, whose square is proportional to the associated light intensity.

Using the 4 normalized vectors $\mathbf{v}_k(l)/\|\mathbf{v}_k(l)\|$ as a basis, the Berreman vector obtained from Eq. (5) can be expressed as:

$$\boldsymbol{\psi}^{2\omega}(l) = \sum_{k=1}^{4} a_k \mathbf{v}_k(l)/\|\mathbf{v}_k(l)\| \qquad (9)$$

where $a_k$ are complex coefficients. Those coefficients can be computed through the scalar products of $\boldsymbol{\psi}^{2\omega}(l)$ and $\mathbf{v}_k(l)$, taking into account that, in general, the 4 vectors $\mathbf{v}_k(l)$ are not orthogonal to each other.

Once written in the form (9) the correction sought for $\boldsymbol{\psi}^{2\omega}(l)$ due to the optic environment can be easily obtained. The corrected vector is simply

$$\boldsymbol{\psi}'^{2\omega}(l) = \sum_{k=1}^{4} a'_k \mathbf{v}_k(l)/\|\mathbf{v}_k(l)\| \qquad (10)$$

where the new coefficients are

$$a'_k = a_k \|\mathbf{v}_k(l)\| \qquad (11)$$

This rescaling of the coefficients using the square root of the intensities ensures an adequate contribution to the electric and magnetic fields of each normal mode of the $2\omega$ wave generated according to the characteristics of the surrounding medium. As has been said before, if the medium has a photonic nature the norms $\|\mathbf{v}_k(l)\|$ can be very different near the bandgaps. Far from the gaps, all of them will be close to unity, but inside the gaps, some of them can be much smaller and, on the contrary, much larger than unity at the resonances at the edges of the forbidden bands. Correction (11) thus accounts for the different degrees of penalty (or enhancement) of each normal mode in the generation of the $2\omega$ field within the material.

Combining (10) and (11) it finally results

$$\mathbf{\psi}'^{2\omega}(l) = \sum_{k=1}^{4} a_k \mathbf{v}_k(l) \qquad (12)$$

4. Once we have the Berreman vector of the $2\omega$ wave generated at each layer, we calculate the total SHG intensity outside the material.

The output Berreman vector $\mathbf{\psi}^{2\omega}_{output}(l)$ of the second-harmonic field due to the contribution of the $l$th layer is obtained by propagating $\mathbf{\psi}'^{2\omega}(l)$ outside the sample, in the form

$$\mathbf{\psi}^{2\omega}_{output}(l) = U^{2\omega}(n)\ldots U^{2\omega}(l+1)\mathbf{\psi}'^{2\omega}(l) \qquad (13)$$

By adding the contributions of all the layers, we have the total SHG field, which will have a Berreman vector

$$\mathbf{\psi}^{2\omega}_{tot,output} = \sum_{l=1}^{n} \mathbf{\psi}^{2\omega}_{output}(l) \qquad (14)$$

The intensity and polarization of the forward SHG light can now be deduced from $\mathbf{\psi}^{2\omega}_{tot,output}$ following standard procedures [15]. The SHG wave in the backward direction can also be obtained by propagating $\mathbf{\psi}^{2\omega}_{tot,output}$ towards the left. The procedure can also be extended to allow incident beams propagating from left to right and from right to left. We have implemented the above calculations on a computer program written in Mathematica. Taking typically $n=1000$ layers, the execution time on a personal computer is only a few seconds.

We will now turn to apply the above procedure to some examples. First, we will check the method in two cases where the solution is known in order to obtain a basic validation of our results. As we previously mentioned, we will take our examples from the world of liquid crystals. More specifically we will use the recently discovered so-called ferroelectric nematic phase as the non-linear medium able to generate the SH field. As will be shown, that phase is very flexible and can adopt a great variety of molecular arrangements of great interest for our purposes. We will treat several structural configurations with different molecular arrangements that will be specified below.

### 3. Study of the SHG performance in the $N_F$ and $N_F^*$ phases.

*- SHG in $N_F$ phase (homogeneous material)*

In ordinary nematic (N) liquid crystalline phases, the constituent molecules are rod shaped, present long-range orientational order and have no positional order. The long axes of the molecules are aligned along a preferential direction called molecular director. Typically, N phases are uniaxial and no spontaneous polarization can appear due to the so-called head-to-tail invariance of the molecular arrangement along the director. However, recently the $N_F$ phase has been discovered [17-27]. In these phases the head-to-tail invariance is broken and spontaneous polarization appears along the director (see Fig. 1). These phases are very promising for nonlinear optical applications since bulky donor-acceptor groups can be incorporated along the long molecular axis, which results in high nonlinear optical responses. Some preliminary studies of the SHG features of $N_F$ liquid crystals have already been done [37-41]. In particular in ref. [41] a complete characterization of the second-order susceptibility tensor has been carried out in the $N_F$ compound RM734 [20]. Two independent components of this tensor are compatible with the symmetry of the phase, assuming Kleinman conditions. The corresponding second susceptibility tensor in contracted notation is:

$$\begin{pmatrix} d_{11} & 0 & 0 & 0 & d_{13} & d_{13} \\ d_{13} & 0 & 0 & 0 & 0 & 0 \\ d_{13} & 0 & 0 & 0 & 0 & 0 \end{pmatrix} \qquad (15)$$

where the *x*-axis is along the molecular director and *y* and *z* are two orthogonal directions in the plane perpendicular to *x*. In ref. [41], it has been stated that the $d_{11}$ coefficient is one order of magnitude higher than $d_{13}$ ($d_{11} = 5.6$ pm/V, $d_{13} < 0.6$ pm/V). In the present calculations we have neglected $d_{13}$, and have assumed a dispersion profile for the ordinary ($n_o$) and extraordinary ($n_e$) refractive indices described by the Cauchy formulas $n_e(\lambda) = 1.8447 + 24151/\lambda^2$, $n_o(\lambda) = 1.6347 + 24151/\lambda^2$ with the wavelength $\lambda$ expressed in nm.

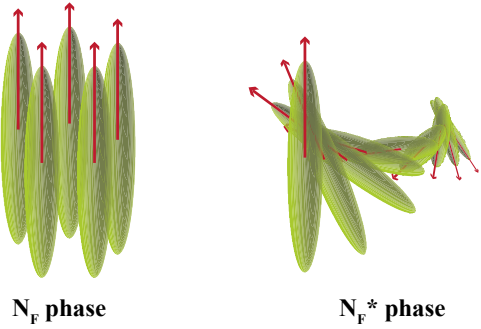

**$N_F$ phase**  **$N_F$* phase**

Fig. 1 Schematic representation of the structure of the $N_F$ and $N_F$* phases. Arrows indicate the direction of the local polarization.

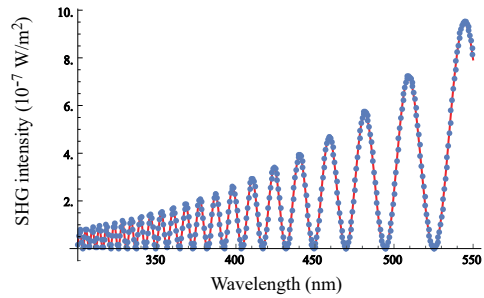

Fig 2. SHG intensity vs wavelength of the SH light calculated by our method (blue dots) and using the expression 16 (red line). The intensity of the input fundamental light is $10^6$ W/m$^2$.

The SHG signal has been calculated for an input fundamental light linearly polarized along *x* and of intensity $10^6$ W/m$^2$. Under these conditions, the SHG light is also *x*-polarized (ee→e conversion). The calculations have been carried out in a sample of 40 μm thickness. For this simple case, there exists an analytical expression for the SHG signal that is given by [2]:

$$I^{2\omega} = I^{\omega 2} \frac{32\pi^2 d_{11}^2}{\lambda^2 \varepsilon_0 c (n_e^\omega)^2 n_e^{2\omega}} \frac{\sin^2 \frac{\Delta k L}{2}}{(\Delta k)^2} \qquad (16)$$

where $I^\omega$ and $I^{2\omega}$ stand for the intensities of the fundamental and SHG lights, respectively and $\Delta k = 4\pi(n_e^{2\omega} - n_e^\omega)/\lambda$, being $\lambda$ the wavelength of the fundamental light in vacuum and $L$ the sample thickness. Dots in Fig. 2 represent the SHG intensity vs. wavelength of the SH wave obtained by using our calculation procedure whereas the red line represents the curve obtained with Eq. (16). As can be seen, an excellent agreement is obtained between both curves. It should be noted that the procedure we have developed gives the SHG intensity in absolute units.

*- SHG in the chiral $N_F$ phase (inhomogeneous material)*

Recently, the chiral version of the $N_F$ phase has also been discovered ($N_F$*) [14,28-31]. The structure is similar to that of the traditional chiral nematic (or cholesteric) liquid crystalline phase N*, i.e., it presents a helical arrangement of the molecules. The main difference is that the $N_F$* phase exhibits spontaneous polarization along the local director (see Fig. 1). Due to this fact, the material is SHG active. Both N* and $N_F$* phases present photonic properties, showing a forbidden band that prevents the propagation of the light in the frequency range inside the band for circularly polarized light with the same handedness as the helix.

The SHG response in liquid crystals with helical molecular arrangements has been addressed in the case of the Smectic C* (SmC*) phases [7,8]. The resolution of this SHG problem is interesting from a basic point of view since the existence of a periodic modulation of the dielectric tensor gives rise to new PM possibilities. In the SmC* case, it is possible to obtain analytic solutions for the eigenstates of both for the fundamental and SH waves that propagate through the material. In ref. [7], an exhaustive analysis of the different possibilities of PM is carried out for light propagating along the helix axis. The theory does not require the SVAA, and provides a numerical solution for the general regime of SHG propagation. Based on this theory, Hoshi et al [8] obtained a general analytical description of the SHG phenomena in the SmC* phase. The theory can be easily adapted to account for the SHG performance of $N_F$* phases by modifying the local dielectric and $d_{ijk}$ tensors according to the new symmetry. We have compared the SHG response obtained in a $N_F$* phase by using our procedure and the analytical theory by Hoshi et al [8]. The comparison is an interesting test of the validity of our calculation method, because the final formulas in ref. [8] are very complicated, especially when the SVAA does not hold, i.e., at the PBGs. For the comparison we have considered a sample of 30$p$ of thickness ($L$=8.27 μm) with a helical pitch $p$=0.2756 μm. We have used the same optical parameters as described in the homogeneous case for the local dielectric tensor and second-order susceptibility. An input fundamental light circularly polarized with the same handedness as the helix, and propagating forward along the helix axis, has been used. Figure 3a represents the reflectance spectrum of the sample. It is worth mentioning that in N* and $N_F$* liquid crystals only one PBG appears, centered at a wavelength $\lambda = \bar{n}p$, being $\bar{n}$ the averaged refractive index. Figs. 3b and c show the forward and backward SHG intensities vs. SH wavelength, respectively, using our procedure (blue dots) and Hoshi's analytical expression (red line). As can be seen, again an excellent agreement is found between both calculations. Two different PMs (A and B) are observed in the spectrum range depicted in the figure. The PM condition holds both for forward and backward propagation of the SHG wave. The intensity corresponding to PM A (343 nm) is dominant in backward propagation. In the case of PM B the wavelength of the PM (532 nm) coincides with that of the long-wavelength photonic edge (LWE) of the band, and the output intensity is the same for both directions of propagation (note the different scales in the ordinate axes in Figs. 3 b and c). We are now going to identify both PM processes.

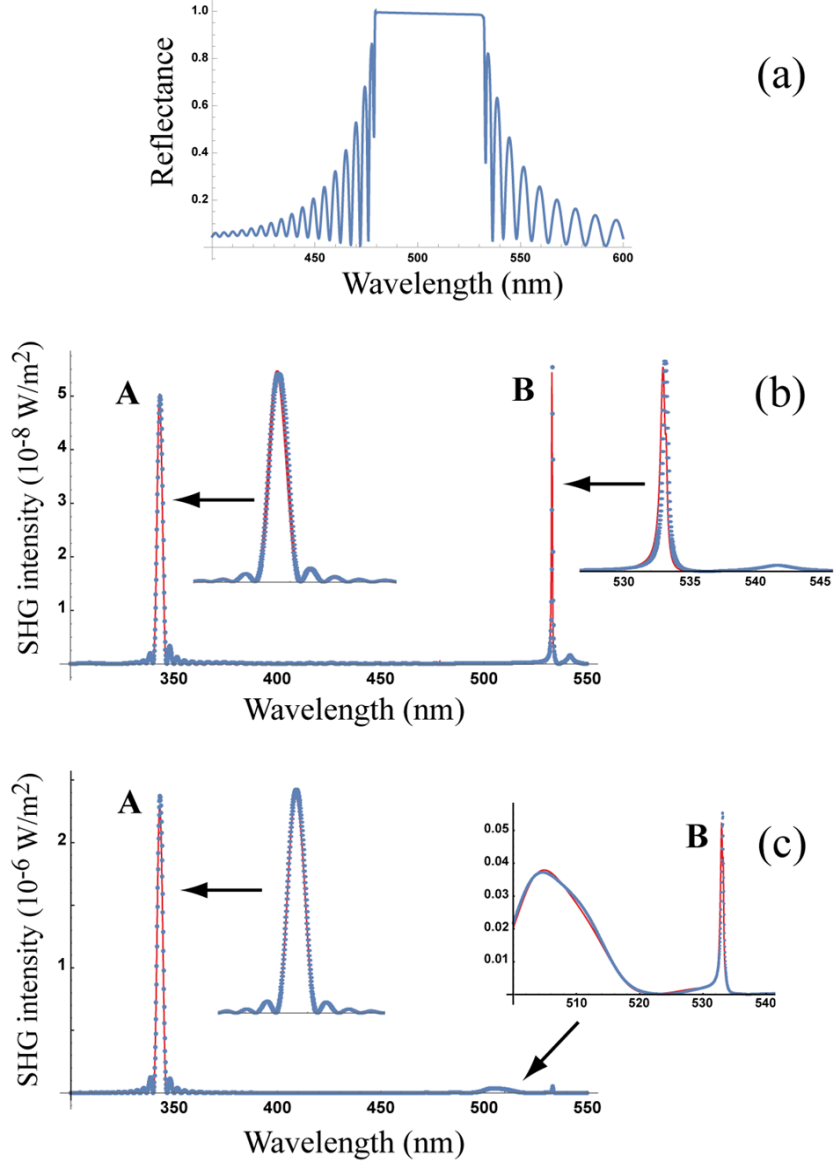

Fig. 3 (a) Reflectance spectrum of the $N_F^*$ sample. Forward (b) and backward (c) propagating SHG intensities vs the SH wavelength. The input light is circularly polarized with the same handedness as the helix, and propagates forward with an intensity of $10^6$ W/m². The blue dots and the red line correspond to our numerical method and Hoshi's analytical expression respectively. In (b) and (c) the different PM profiles (A and B) have been enlarged.

To do this identification we must recall that the dispersion relation of the light modes of frequency $\omega$ in a N* phase is given by [42]:

$$l = \pm \left[ (k_0^2 \bar{n}^2 + q^2) \pm \sqrt{4 k_0^2 \bar{n}^2 q^2 + a^4 k_0^4} \right]^{1/2} \qquad (17)$$

where $l$ is the wave vector, $q = 2\pi/p$, $k_0 = \omega/c$, $\bar{n}^2 = \frac{1}{2}(n_e^2 + n_o^2)$, $a^2 = \frac{1}{2}(n_e^2 - n_o^2)$. A schematic representation of the dispersion relationship is depicted in Fig. 4. The branches with positive (negative) slope represent forward (backward) propagating waves. The eigenmodes are essentially circularly polarized except at the edges of the gap or for very high frequencies. At the low (high)-frequency edge of the gap the $l=0$ mode is a stationary wave whose electric field locally oscillates in a direction parallel (perpendicular) to the molecular director [42]. Blue branches correspond to polarizations with the same handedness as the helix whereas red branches stand for the opposite.

We can now easily identify the PMs depicted in Fig. 3 as combinations of modes in the dispersion curve. The PM A in Fig. 3c (343 nm) corresponds to the combination denoted as A in Fig. 4 (green arrow). In our case, the input light is a circularly polarized mode travelling forward, giving rise to a counter-propagating SHG light of the same handedness. Since the PM condition can be written as $l^{2\omega} = 2 l^{\omega}$, Eq. (17) implies the following relation [7]:

$$\frac{2\omega}{c}[\bar{n}(2\omega) + \bar{n}(\omega)] \approx 3q \qquad (18)$$

which is, indeed, satisfied for $\lambda/2 = 343$ nm ($\bar{n}(343$ nm$) = 1.945$ and $\bar{n}(646$ nm$) = 1.791$ according to the Cauchy formula). It can be shown that the dependence of the SHG intensity scales as $L^2$. It is worth mentioning, that a much smaller intensity is observed in forward propagation (left peak in Fig. 3b). This SH light arises by the residual fundamental light always reflected inside the sample, that gives rise to a PM configuration equivalent to A in Fig. 4, but with the waves involved in the process propagating in the opposite directions.

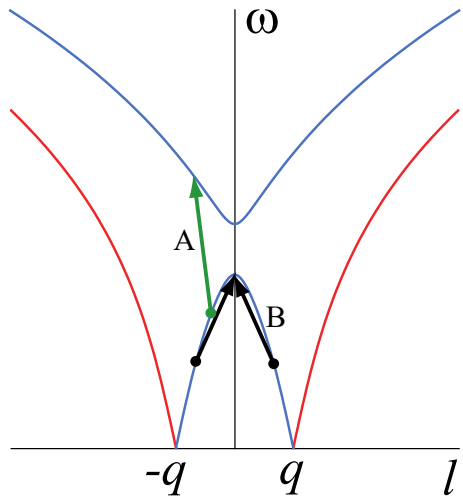

**Fig. 4** PM combinations schematized on the dispersion curves of the optical eigenmodes of the $N_F^*$ phase. Points of the branches with positive (negative) slope represent modes propagating forward (backward). Except at the band edges the polarization of all modes is essentially circular, with the same handedness as the helix for blue branches and with opposite handedness for red branches. In the PM process denoted by A, two photons of frequency $\omega$ on the origin of the green arrow, give rise to a photon of frequency $2\omega$ on the point indicated by the arrowhead. In the B process, the two photons on the origins of the black arrows, both of frequency $\omega$ and with opposite wave vectors $l$, combine to give a $2\omega$ photon with wave vector $l=0$ at the LWE of the gap (head of the black arrows).

Finally, the PM B (right peak in Figs. 3b, c) corresponds to the mode combination depicted as B in Fig 4 (black arrows). The PM condition takes place at the LWE of the gap and occurs for two counter-propagating fundamental input lights. The PM condition can be written as [7]:

$$\frac{2\omega}{c}n_e(2\omega) = q \qquad (19)$$

which is satisfied for $\lambda/2$=532 nm ($n_e(532 \text{ nm}) = 1.930$).

This kind of PM gives rise to very high performances when the sample is illuminated from both left and right sides simultaneously, so that the two counter-propagating input lights present similar intensities. In fact, it can be shown that the SHG intensity scales as $L^4$. However, in our case, the fundamental light travelling backward is only the residual counter-propagating mode due to the small reflection that necessarily occurs inside the modulated sample. Note that the reflectivity is very small for the wavelength of fundamental light ($\lambda \approx 1064$ nm) but it is, nevertheless, high enough to give rise to an important SHG intensity.

It is interesting to note there is no a PM similar to B at the short-wavelength edge (SWE) of the band (480 nm, see Fig. 3a). This absence can be understood if we recall that the SH electric field for this mode would oscillate in a direction perpendicular to the molecular director and, therefore, according to the proposed form for the $d_{ijk}$ tensor, there is no possibility for SHG.

The PMs discussed here are just two examples of a much larger variety of PM combinations existing in this kind of structures [7]. In fact, we have checked that more PMs are obtained, if a larger spectral range is examined or other input polarizations are used, although the SHG intensities are in general smaller. In all cases we have found an excellent agreement between both calculation procedures which, given the complexity of this case, gives us full confidence in the validity of our method.

### - SHG in the $N_F^*$ phase under electric field

Now we calculate the SHG response of a sample with a more complicated modulation. We will consider again the $N_F^*$ sample but now subjected to an electric field perpendicular to the helix. The linear optical properties in this configuration have been previously studied in refs [28-31]. For this example, there are not available procedures to calculate the SHG intensity so far. However, the versatility of our calculation method allows us to address this problem.

Firstly, we will briefly describe the structure of the $N_F^*$ phase under electric field. In a classical nematic material, due to the head-to-tail invariance, the true periodicity of the structure is $p/2$. This fact gives rise to the appearance of just a single PBG centered at a wavelength $\lambda = 2\bar{n}(p/2)$. In contrast, in the case of a $N_F^*$ phase, due to the local spontaneous polarization, the structure periodicity is $p$. However, optically, both structures are equivalent and the reflectance spectrum presents a unique PBG centered at $\lambda = \bar{n}p$ (half pitch band). The differences arise when the material is subjected to an electric field perpendicular to the helix axis. In this case, the coupling of the electric field with the local polarization gives rise to a distortion in the helix profile. Then, the resulting structure presents a reflectance spectrum with multiple PBGs appearing at wavelengths $\lambda = 2p\bar{n}/m$ with $m = 1,2,3...$ In ref. [31] a detailed calculation of the molecular director profile versus electric field has been performed and a study of the photonic properties of the structure has been carried out. We have simulated the structure of the $N_F^*$ phase, distorted under electric field, following the procedure explained in that reference. Briefly, if we denote $\phi(z)$ the azimuth angle of the molecular director with respect to the corresponding at $z = 0$, the profile of this angle can be calculated by solving the equation [28]:

$$K_2 \frac{d^2\phi}{dz^2} = PE\sin\phi \qquad (20)$$

where $K_2$ is the twist elastic constant, $P$ the polarization and $E$ the electric field. In this equation the so-called dielectric and flexoelectric couplings with the electric field have been neglected. The parameters used in the calculations are $K_2 = 6.5$ pN and $E = 8$ V/mm, being the rest of the parameters required for the simulation the same as in the case of the $N_F^*$ phase under no field.

Figure 5 represents the azimuth angle $\phi/(2\pi)$ versus $z$. The distortion of the molecular azimuth profile induced by the electric field can be observed as a slight undulation of the curve $\phi$ vs $z$. As can be seen, even for a small electric field, an appreciable distortion can be observed. For higher electric fields, the distortion becomes more evident and the helical pitch slightly increases. If the field is above a certain threshold the helix is suppressed.

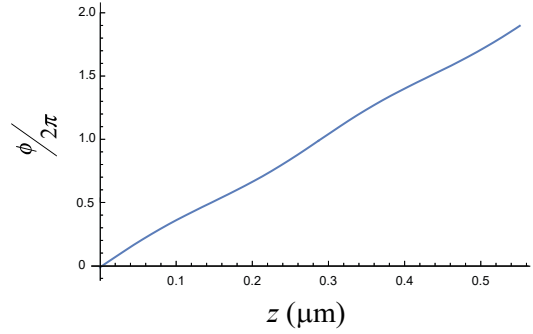

Fig. 5  $\phi/(2\pi)$ versus z for the $N_F^*$ phase under an electric field of 8 V/mm, perpendicular to the helix.

Fig 6a shows the reflectance spectra for a fundamental light circularly polarized with the same handedness as that of the helix at normal incidence. As can be seen, three PBGs can be observed in the depicted spectral range. They correspond to the different harmonics of the fundamental reflection band (1 in Fig. 6a), and are related to the different Fourier components of the helix distortion. This spectral configuration is very interesting for SHG purposes since the photonic materials with multiple PBGs can present high performance when both the fundamental and SH lights are enhanced by resonance simultaneously.

Although the new structure presents a complex SHG response due to the rich variety of PBGs that exhibits, we will focus on the SHG processes in the vicinity of the half-pitch PBG (2 in Fig. 6a). Figs. 6b and c represent the SHG intensity in forward and backward directions, respectively. As can be seen, in both cases two different PMs can be observed, corresponding to two the edges of the band. The highest intensity occurs for the LWE of the band (532 nm). Remarkably, the small distortion induced by the electric field multiplies by three the SHG signal under no field. Contrary to that case, here a SHG peak also appears at the SWE of the band (480 nm). Both PMs scale as $L^4$.

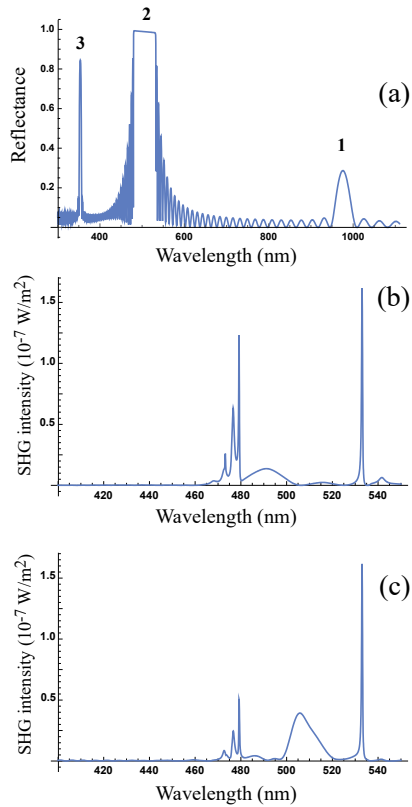

Fig. 6 (a) Reflectance spectrum of the $N_F^*$ sample under an electric field of 8 V/mm. The full-pitch, half-pitch, and third-harmonic bands are denoted by 1, 2, and 3, respectively. (b) Forward SHG intensity vs wavelength of the SH light. (c) The equivalent case to (b) but for the SH light travelling backward. The intensity of the input fundamental light is $10^6$ W/m$^2$.

We will try to qualitatively interpret the obtained results. As previously mentioned when studying the SHG performance under no electric field, a very efficient PM at the PBG edge appears if there are two fundamental light beams in the sample propagating in opposite directions. In our case, the fundamental input light propagates essentially forward but a counter-propagating residual light with the same polarization always appears inside the modulated material. This residual contribution becomes more relevant under field, since the fundamental light is now in the vicinity of the full pitch PBG (1 in Fig 6a), which results in a more intense reflection of the fundamental light. Obviously, the most efficient SHG conversion would happen when both the fundamental and SHG lights coincide with the edges of the PBGs simultaneously, but this situation requires a particularly favorable index dispersion law and it is impossible to attain in the present case.

Finally, we give a brief account about the appearance of the PM at the SWE of the gap (left peak in Figs. 6b and c). In this case, it can be shown that the polarization of the corresponding $2\omega$ light mode is not perfectly perpendicular to the local director. Thus, a small SHG response is possible through the $d_{11}$ coefficient.

## 4. Conclusions

We have presented a numerical procedure for calculating the SHG field generated by an anisotropic material with a 1D modulation of the dielectric permittivity. The method can be applied for arbitrary modulations and even when the input light has an oblique incidence. The method is based on the Berreman 4x4 matrix formalism and, in fact, it is a generalization of that procedure to incorporate SHG processes. The strategy results to be a powerful tool to explore the SHG performance of complex optical systems. Compared to existing models, the present approach represents a significant step forward in the study of the SHG response of nonlinear anisotropic inhomogeneous media.

As an example of application of the method, the SHG response of a $N_F*$ liquid crystal, subjected to electric fields, has been studied. The obtained results show attractive potentialities of these materials for developing cheap and high-performance optical devices. In the near future we are planning to explore experimentally the results predicted in this work.

**Funding**